# Structural similarity enhances interaction propensity of proteins.

D. B. Lukatsky[*], B. E. Shakhnovich[**], J. Mintseris[**], and E. I. Shakhnovich[*]

[*]*Department of Chemistry and Chemical Biology, Harvard University, Cambridge, MA 02138*

[**]*Bioinformatics Program, Boston University, Boston, MA 02215*

Corresponding author: Prof. Eugene Shakhnovich, *Department of Chemistry and Chemical Biology, Harvard University, 12 Oxford St., Cambridge MA 02138.*
Email: eugene@belok.harvard.edu
Ph.:   617-495-4130
Fax:   617-384-9228

We study statistical properties of interacting protein-like surfaces and predict two strong, related effects: (i) statistically enhanced self-attraction of proteins; (ii) statistically enhanced attraction of proteins with similar structures. The effects originate in the fact that the probability to find a pattern self-match between two *identical*, even randomly organized interacting protein surfaces is always higher compared with the probability for a pattern match between two *different*, promiscuous protein surfaces. This theoretical finding explains statistical prevalence of homodimers in protein-protein interaction networks reported earlier. Further, our findings are confirmed by the analysis of curated database of protein complexes that showed highly statistically significant overrepresentation of dimers formed by structurally similar proteins with highly divergent sequences ("superfamily heterodimers"). We predict that significant fraction of heterodimers evolved from homodimers with the negative design evolutionary pressure applied against promiscuous homodimer formation. This is achieved through the formation of highly specific contacts formed by charged residues as demonstrated both in model and real superfamily heterodimers.

# Introduction

Several independent analyses of accumulating high-throughput and specific data on protein-protein interactions (PPI) revealed a general statistical bias for homodimeric complexes. In particular, PPI networks from four eukaryotic organisms (baker's yeast *S.cerevisiae*, nematode worm *C.elegans*, the fruitfly *D.melanogaster* and human *H.sapiens*) obtained from high-throughput experiments contain 25-200 times more homodimeric proteins than could be expected randomly[1]. The same trend was observed in detailed analysis of confirmed protein-protein interactions - a phenomenon called "molecular narcissism" (S. Teichmann, private communication). It was also shown experimentally[2] that the sequence similarity is a major factor in enhancing the propensity of proteins to aggregate. The physical or evolutionary basis for these striking observations remains unexplained.

Here, we propose a simple model of protein-protein interactions and show that the observed preference for homodimeric complexes is a consequence of general property of protein-like surfaces to have, statistically, a higher affinity for self-attraction, as compared with propensity for attraction between *different* proteins. Moreover, we predict that the same effect of statistically enhanced attraction is operational for protein pairs of *similar structure*, even in the case when their aminoacid sequences are far diverged.

The predicted physical effect of statistically enhanced attraction of structurally similar proteins has significant implication for evolution of protein-protein interactions: It suggests a duplication-divergence route by which many modern protein complexes could have evolved from earlier homodimers through sequence, divergence of paralogous genes under the constraint of keeping structures less divergent.

## Model: statistically enhanced self-attraction of proteins

We begin with a residue-based model of a protein interface[3], Figure 1A (see Methods). This model allows for all twenty aminoacid types to be represented as hard spheres and randomly distributed on a planar, circular interface. Multiple surfaces are generated whereby amino acids are placed randomly and their identities are drawn randomly from a probability distribution corresponding to aminoacid composition on real protein surfaces (see Methods), and we impose that the total number of residues, $N$, in each surface is fixed. All chosen parameters correspond to a typical protein interface[4; 5; 6; 7] (Methods). Using this model, we investigated the statistical interaction properties of such random surface pairs. Residues of two interacting surfaces (IS) interact via the Miyazawa-Jernigan (MJ) residue-residue potentials[8], and we assume that two residues are in contact if they are separated by the distance less than $8 \overset{\circ}{A}$. For each realization of two surfaces, we fixed the inter-surface separation to be $\sim 5 \overset{\circ}{A}$. We then proceeded to rotate one surface with respect to the other, to find the lowest interaction energy for each pair. This way we obtained the Lowest Energy Distribution (LED) of the inter-surface interaction energies for different random realizations of IS.

The first task is to compare random heterodimers (superimposed pairs of *different*, randomly chosen random surfaces) and homodimers (self-superimposed surfaces, *i.e.* each surface is superimposed with a reflected image of itself). The results of these calculations [for a specific, average aminoacid composition from a homodimer dataset[6]] are shown in Figure 1C. The key result is that random model protein surfaces have *always* a statistically higher propensity for self-attraction as compared with random heterodimers. The tail of LED for homodimers is *always* shifted towards lower energies with respect to random heterodimers.

The simple physical reason for this key finding can be illustrated using a toy model, Figure 2, where hydrophobic residues are randomly distributed on a flat lattice surface with $N_0$ sites. For the sake of simplicity, we consider only strong binding in both cases (homo- and heterodimers) where all hydrophobic residues on both interacting surfaces match. First, we estimate the probability, $p_{\text{homo}}$, that a given random pattern (with a fixed number of residues $N$) forms a strongly bound homodimer (self-matching pattern). Such a self-matching pattern can be obtained by distributing $N/2$ hydrophobic residues at random, by selecting an *arbitrary* axis of symmetry, and finally, by distributing the remaining $N/2$ residues at symmetrically reflected positions with respect to this axis. Therefore, $p_{\text{homo}}$ is simply $\simeq Q_{\text{self}}/Q$, where $Q \simeq N_0!/N!(N_0-N)!$ is the total number of distinct patterns with $N$ hydrophobic residues and $Q_{\text{self}} \simeq (N_0/2)!/[(N/2)!(N_0/2-N/2)!]$ is the number of self-matching patterns. Now we can compare $p_{\text{homo}}$ with an analogous probability, $p_{\text{hetero}}$, for a pair of *distinct* surfaces to form a strongly bound heterodimer. It is easy to see that only one other pattern will form a perfect complement to a non self-matching pattern, therefore $p_{\text{hetero}} \simeq 1/Q$. Thus, the ratio, $p_{\text{homo}}/p_{\text{hetero}}$, is a large number of the order of $Q_{\text{self}}$. Intuitively, this effect arises simply because in order to obtain a strongly bound homodimer, one needs to match (by random sampling) only $N/2$ hydrophobic contacts, while all $N$ contacts need to be matched for a strongly bound heterodimer. Therefore, although locations of residues on each surface are disordered, it is more likely to find two identical, random surfaces that strongly attract each other, as compared with two different random surfaces because it is more probable to symmetrically match a half of a random pattern with itself than with a different random pattern, which requires a full match.

This consideration suggests that patterns constituting strongly bound homodimers are more symmetrical than an average random pattern. We probed the symmetry of such patterns (selected from the low-energy tail of the homodimeric *P(E)*) computing the correlation function of aminoacid density, and confirmed this prediction, Figure 8 (Methods). We emphasize that the predicted effect holds for *any* aminoacid composition and for *any* type of the interaction potential between amino acids (an analytical theory[9] that further develops simple ideas presented here confirms the universality of the effect).

These results suggest, most importantly, that homodimers were selected with a higher probability (than would be expected randomly) in the course of evolution as functional protein-network motifs. The energy difference between the maxima of *P(E)* for homo- and heterodimers, Figure 1C, provides an estimate for the strength of the predicted effect, $\sim 0.1 k_B T \simeq 60 \, \text{cal/mol}$ enthalpy reduction (on average) per one homodimeric interface residue, where $k_B$ is the Boltzmann constant and $T$ is the temperature. Assuming that there are 50 interface residues on average, per each protein complex, we predict that homodimers should occur with the probability of $\exp(5) \simeq 150$ times higher than it would be expected simply based on the average protein concentrations, and without taking into account the predicted effect. This provides a possible explanation for the observed, anomalously high frequency of homodimers (25-200 times higher than expected) in protein interaction networks[1].

*Simple model for evolutionary selection of strongly interacting homodimers*

Our results imply that homodimers occur with higher probability (than would be expected randomly) in the "soup" of randomly exposed protein surfaces. Correspondingly they could be preferentially selected in the course early of evolution, as functional protein-network motifs. This scenario, which we call "one-shot selection", can be modeled in our model and tested by comparing aminoacid compositions in selected model strongly interacting homodimers with interfacial aminoacid composition of real homodimers.

To this end we selected strongly self-interacting surfaces (*e.g.*, with the interaction energy $E$<-3.3, Figure 1C) from the set of all randomly generated ones. Next we checked the aminoacid compositions of these selected, strongly interacting homodimeric surfaces and compared it with the observed compositions in homodimeric interfaces of proteins[4; 6]. The resulting aminoacid composition of the selected, strongly attracting homodimeric interacting surfaces is presented in Figure 3 in terms of the interface propensity for each of 20 residues, where the model interface propensity is $\ln(f_\alpha / f_\alpha^0)$, with $f_\alpha$ and $f_\alpha^0$ being the fraction of residue type $\alpha$ in the *selected* set of surfaces and the average fraction of residue $\alpha$, in *all* protein surfaces (which coincides with probability distribution with which we selected aminoacid types to generate random surfaces), respectively. We emphasize that $f_\alpha^0$ is the input to the model from experimental data, and $f_\alpha$ is produced by the model. The model results correlate with the observed experimental interface propensities[6] with the correlation coefficient $R \simeq 0.93$, Figure 3. Such a strong correlation between the model and experiment provides a consistency test for the model. Indeed if we change selection criterion from that of low-energy self-interacting surfaces to a "window" of higher interaction energies we observe a sharp transition in aminoacid composition of surfaces selected in a sliding window of interaction energies, from the highly correlated with experiment value of +0.93 (when strongly interacting surfaces corresponding to the left tail of the homodimeric LED on are selected) to the anti-correlated with experiment value of –0.91 (when mutually repulsive homodimeric surfaces at the right tale of the LED are selected).

We emphasize that high correlation between the model predictions and experimental data, Figure 3, is much more than just a correct yet trivial prediction for the *relative* propensity of hydrophobic and hydrophilic residues at protein homodimeric interfaces. The model correctly predicts the relative propensities *within* the hydrophobic and hydrophilic groups of amino acids. This is demonstrated in the reshuffling control calculation (see Methods), which results in the highly statistically significant *p*-value of $p \simeq 0.00006$. Finally, we stress that the predicted high correlation between the model and experiment is robust with respect to the choice of the effective, residue-residue potential. The model calculation with *e.g.* the Mirny-Shakhnovich potential[10] (data not shown) leads to the high correlation coefficient, $R=0.91$, between the model and experimental results.

**Structural similarity enhances interaction propensity of proteins**

We now turn to the second key finding of this paper – the prediction of the enhanced attraction between *structurally similar* protein pairs, even in the case when their aminoacid sequences have low sequence identity. Such proteins with high structural similarity and low sequence identity (usually below 25%) are commonly classified as belonging to a particular protein *superfamily* (see *e.g.*, Ref.[11]). Correspondingly, we term interacting pairs of such structurally similar proteins as *superfamily heterodimers*. The two interacting surfaces of a model superfamily heterodimer are represented in Figure 1A and B. The *spatial positions* of amino acids within these two surfaces are identical, however, the aminoacid identities (colours in Figure 1A and B) are randomly reshuffled. We have computed the LED for the interaction energies, *P(E)*. The results of these calculations are shown in Figure 1C. The key message here is that again, the probability distributions *P(E)* for superfamily heterodimers are systematically shifted towards lower energies, as compared with the corresponding *P(E)* for random heterodimers. Similar to the case of homodimers, for superfamily heterodimers this effect holds for any aminoacid composition and for any type of the inter-residue interaction potential.

The principal question remains whether the predicted effect of statistically enhanced attraction between structurally similar proteins is observed in real protein complexes. To answer this question we analysed structural similarity of interacting protein chains using the literature-curated, non-redundant dataset of two different types of crystallized heterodimeric complexes (see Methods). These are 115 *obligate* and 212 *transient* complexes[12]. While obligate complexes are biologically functional only as permanent assemblies, each chain in a transient complex can function on its own. Therefore, obligate complexes are stronger bound (on average) than transient complexes, and our model predicts that the interacting chains in obligate complexes should be structurally more similar (on average) as compared with transient complexes. The structural similarity of two proteins can be quantitatively characterized by the Z-score[13], where higher Z scores indicate greater similarity[13; 14]. The key finding here is that the fraction of interacting chains with high structural similarity involved in obligate complexes is strikingly larger than the fraction of chains with high structural similarity involved in transient complexes, which in turn is much greater than random control where chains constituting control "heterodimers" are selected at random from pdb, Figure 4. This is the key finding of our paper. This result is highly statistically significant. For example, at $Z>2$, the absolute difference between the frequencies of obligate and transient complexes constitutes 20%, that is about 28 standard deviations of the control frequency. Higher $Z$ cutoffs yield even larger observed differences, up to 130 standard deviations at $Z>10$. We emphasize again that *all* protein complexes selected for the structural similarity analysis have very low values of sequence identity between interacting chains, below 25%.

The enhanced *structural* similarity of superfamily heterodimers leads, statistically, to a larger number of *favourable*, inter-surface contacts, which in turn, enhances statistically the interaction propensity of such structurally similar proteins. The statistics of interface contacts in both real and model protein complexes is shown in Figure 5A and Figure 5B, respectively. The key observation here is that the frequency distributions for the number of contacts in homodimers and obligate heterodimers are shifted towards the *larger* number of contacts (per one interface atom) as compared with transient heterodimers, Figure 5A. This result is highly statistically significant, as the computed Kolmogorov-Smirnov values (comparing the distributions) demonstrate, Figure 5A. The model results, Figure 5B, demonstrate a qualitatively similar trend. Remarkably, and also consistent with the model results, the distributions for the number of contacts in obligate heterodimers and homodimers are very similar (large *p*-value). Therefore, both effects - the statistically enhanced self-attraction of proteins and the enhanced attraction between structurally similar proteins have primarily a structural origin.

**Divergent evolution of homodimers and negative design**

Our finding of increased frequency of superfamily heterodimers in obligate and transient complexes opens a possibility that a significant fraction of superfamily heterodimers evolved from homodimers. An example illustrating a divergent nature of superfamily heterodimers can be seen using a phylogenetic analysis of a prokaryotic DNA bending protein complex (1ihf), Figure 6 (see Methods), where paralogous genes constituting monomers of an obligate dimer that apparently originated from the common root, exhibit the degree of sequence divergence in a broad range, from sequence ID close to 100% to as low as 13%.

Since homodimers are (statistically) stronger bound complexes than both random and superfamily heterodimers, the key, physical issue that evolution had to resolve for heterodimers evolving from homodimers is how to select against promiscuous homodimerization. There is a simple, physical common-sense solution to this problem – to place charged residues of opposite signs on the interacting surfaces of superfamily heterodimers. Therefore we test whether our model and Nature use this common-sense solution in evolutionary selection of superfamily heterodimers. Here we only considered the evolution of homodimers towards superfamily heterodimers, *i.e.*, the scenario where the *structures* of evolving protein pairs remain intact, and only their sequences diverge. This represents the most relevant case, where competitive, promiscuous homodimeric interactions are the strongest.

We performed a stochastic design procedure (see Methods) to mimic the evolutionary transformation of homodimers towards heterodimers. This procedure started from the selected, strongly interacting homodimeric surfaces and proceeded to evolve them to strongly interacting superfamily heterodimeric surfaces. In addition to the requirement of strong interaction between surfaces, we also applied a negative design requirement against promiscuous homodimer formation, as a criterion to accept or reject mutations. We compared the resulting frequencies of charge contacts across the evolved superfamily heterodimeric and homodimeric interfaces. We also analysed the corresponding frequency differences of charge contacts in real homodimeric and superfamily heterodimeric complexes. The principal message emerging from this analysis, Figure 7, is that in both real and model protein interfaces there are significantly more favourable $(+-)$ contacts in superfamily heterodimers, as compared with homodimers. The statistical significance of this result is apparent from the analysis of the corresponding frequencies of unfavourable contacts, Figure 7. The interface charges therefore not only provide the positive design for heterodimeric interactions, but also simultaneously protect heterodimers against promiscuous homodimer formation. In particular, specific residues making salt bridges, *e.g.*, Lys-Glu or Arg-Asp and stabilizing heterodimers, at the same time provide the negative design against homodimers, forming unfavourable, similarly charged contacts, such as *e.g.*, Lys-Lys, or Glu-Glu. This finding is in agreement with other investigations of the effect of negative design and the stabilization of protein domains against aggregation [15; 16].

**Discussion and conclusion**

Our model description of protein-protein interactions is highly simplified, yet we suggest that the mechanism for enhanced attraction of structurally similar or identical proteins described here is quite general. The structural similarity graph, Figure 4, that represents the main experimental support for our prediction, is likely to be the rule rather than the exception. Similar statistical trend for enhanced structural similarity of strongly interacting proteins should be observable in larger scale PPI sets, such as the organismal PPI networks[17; 18].

We emphasize that the predicted effect is *statistical* in its nature - we predict a generic law for statistical probability distributions. This law is thus applicable to protein sets rather than to individual proteins. The estimated *average* strength of the effect is as large as few kcal/mol enthalpy gain per one typical homodimeric or superfamily heterodimeric protein complex (with a few tens of interface residues, on average), as compared with heterodimeric complex (with a similar average number of interface residues and similar aminoacid compositions). This is a strong effect, and the predicted enthalpy gain is comparable with the average free energy scale of protein stability. This estimate explains quantitatively the observed overrepresentation of homodimers in protein interaction networks[1].

Further, our findings explain the recent experimental discovery[2] that sequence divergence is a major evolutionary mechanism inhibiting protein aggregation and amyloid formation. It was demonstrated in Ref.[2] that in two large, multidomain protein superfamilies (immunoglobulin and fibronectin type III) of the adjacent domain pairs in the same proteins, more than two-thirds have less than 30% identity. Moreover, only about 10% of the adjacent domain pairs have more than 40% identity. Our prediction of statistically enhanced self-attraction of proteins rationalizes the aggregation mechanism discovered by Dobson *et al.*[2] as evolutionary emerged from the statistically enhanced correlations between *identical* protein interfaces as compared with *different*, promiscuous interfaces.

Our analysis predicted and experimental data confirmed the negative design mechanism against promiscuous homodimer formation. The mechanism that Nature utilizes is simple and intuitive – the selectivity of charge-charge interactions is employed to select against promiscuous homodimers, and at the same time, to increase the stability of heterodimers. The importance of electrostatic interactions in protein-protein recognition has been acknowledged in the literature[15; 16]. The common opinion is that electrostatic interactions confer strong binding at interfaces (positive design). While not inconsistent with our findings this conjecture underestimates another, perhaps even more important role of charge interactions: to confer specificity against a particularly challenging type of promiscuous interactions. A good evidence to support this view is the observed conservation of specific charged residues on protein interfaces[19]. Thus we conclude that while hydrophobic interactions are mainly responsible for tight binding in protein complexes, charged pairs confer specificity and protection against promiscuous homodimeric interactions.

Finally, it is tempting to speculate that enhanced propensity for homodimer and superfamily heterodimer formation allows more duplication events to lead to biologically functional complexes. In turn, this may increase the fitness of the population and provide sufficient evolutionary pressure to fix gene duplications through increasing the phenotypic diversity of mutants. The physical mechanism of protein-protein interactions presented here and its experimental verification are striking examples of how evolution operates within constraints imposed by fundamental physical laws.

## Methods

### Generation of model surfaces

Model protein surfaces are generated by randomly distributing amino acids (20 types, represented by impenetrable hard-spheres with the diameter, $d_0 = 5$ Å) on planar, circular surfaces with the diameter, $D = 70$ Å. In the model calculations the number of amino acids on each surface is fixed, $N=70$, and thus the surface fraction of residues is $\phi = Nd_0^2/D^2 \simeq 0.357$. The chosen parameters correspond to a typical protein interface[4; 5; 6; 7]. The *spatial* positions of amino acids are random, however amino acids are not allowed to inter-penetrate each other, therefore, the minimal, possible separation between any two amino acids can not be smaller than $5$ Å. The *identities* of amino acids are drawn randomly from a probability distribution that specifies the *average* fraction (composition) of each (out of all 20) aminoacid types. *Therefore, the aminoacid composition of a randomly generated surface may differ significantly from the input, average composition.* After both the aminoacid locations and identities for a given surface are generated, this configuration is fixed (quenched).

In order to find the Lowest Energy Distribution (LED) for random *heterodimers*, we generate pairs of random surfaces as described above, superimpose these surfaces in exactly parallel configuration (where surfaces are separated by the distance $\sim 5$ Å), and mutually rotate the surfaces until the minimum of the *inter*-surface interaction energy is found. To compute the LED for model *homodimers*, we superimpose pairs of *identical* random surfaces and repeat the described procedure to find the LED for such pairs. We emphasize that in the latter case, we superimpose each surface with the reflected image of itself - exactly in the way real, *identical* protein pairs would superimpose their surfaces upon interaction and homodimer interface formation.

**Obligate and transient complexes**

The complete lists of all 115 obligate and 212 transient complexes can be found at http://zlab.bu.edu/julianm/MintserisWengPNAS05.html. The details about these datasets can be found in Refs.[12; 20]. In computing the structure similarity distribution, Figure 3, from the entire datasets of complexes, we selected only those complexes, where the interacting protein pairs have sequence identity of less that 25%. The resulting set of high *Z*-score obligate and transient complexes is presented in Table 1. We term such structurally similar heterodimeric complexes (with far diverged sequences of interacting chains) as "superfamily heterodimers".

**Structural similarity of proteins and *Z*-score**

The FSSP database, based on the DALI structure comparison algorithm[13; 21], defines a quantitative measure of structural similarity, the *Z*-score. We used the DaliLite program[13], http://www.ebi.ac.uk/DaliLite/ to compute the *Z*-scores. Only complexes with sequence identities of interacting chains of less than 25% (*i.e.* only superfamily heterodimers) were considered. The control bars in Figure 3 of the paper were computed by picking *random* pairs from *all* 3313 protein DALI domains constituting the Protein

Domain Universe Graph (PDUG)[14], computing their Z-scores, and finally computing the corresponding frequencies of the occurrence of high Z-score PDUG protein domain pairs.

**Positional correlations between amino acids in strongly and weakly bound model homodimers**

We computed the local aminoacid density-density correlation function, $g(\rho)$, for the model surfaces selected as *strongly* interacting homodimers (*i.e.* selected from the left tail of the homodimeric LED, *P(E)*), and compared it with $g(\rho)$ for the *weakly* interacting homodimers (selected from the right tail of *P(E)*). $g(\rho)$ is defined as the probability distribution to find two amino acids (randomly selected within a surface, irrespectively to their identity) to be separated by the distance $\rho$. The results are shown in Figure 8. The key message here is that amino acids in the strongly bound homodimers have higher *positional* correlations as compared with the case of weakly bound homodimers.

**Reshuffling control for model homodimer aminoacid propensities**

We computed the probability distribution function of the linear correlation coefficient, *R,* upon the *partial* reshuffling the identities of residues in the model data set, *i.e.* upon reshuffling separately within the mostly hydrophobic [Cys Met Phe Ile Leu Val Trp Tyr Ala] and mostly hydrophilic [Gly Thr Ser Asn Gln Asp Glu His Arg Lys Pro] groups of residues. This procedure shows negligibly small probability $p(R > 0.93) \simeq 0.00006$ to find the predicted correlation coefficient "by chance" (even assuming a correct redistribution of hydrophobic and hydrophilic residue groups). The complete reshuffling of residue identities leads, of course, to a symmetrically distributed around zero probability distribution with a zero (up to the computer precision) probability of obtaining $p(R > 0.93)$ "by chance".

**Example: Phylogenetic analysis of DNA binding complex**

The implication from our observation of increased frequency of superfamily heterodimers in obligate and transient complexes is that these chains might share an evolutionary relationship, presumably originating from a duplication event yielding homodimeric paralogs. An example of this phenomenon can be seen using a prokaryotic DNA bending protein complex (1ihf). First, we map the orthologs of both chains in the complex on the bacterial clade of the phylogenetic tree. We observe a range of divergence between the two chains in different species ranging in sequence similarity from 13 to 98% (Figure 6). This observation can be interpreted using two parsimonious scenarios. The two chains in species with high sequence similarity have either recently duplicated or have been subject to strong selection since the duplication event.

**Stochastic design procedure**

The stochastic design procedure with the conserved aminoacid compositions attempts a mutation by randomly swapping the identities of a randomly chosen pairs of residues within each of the two interacting surfaces. The attempted mutation is accepted with the standard Metropolis criterion[22] on the lowest (with respect to rotation) value of the inter-protein interaction energy. The lowest value of the inter-protein interaction energy is computed in each MC step. The negative design on homodimer formation is implemented in the MC procedure using the total inter-protein energy in the form $E_{tot} = E_{hetero} - \alpha E_{homo}$, where $E_{hetero}$ and $E_{homo}$ are the interaction energy of the corresponding hetero- and homodimer, respectively, and the strength of the negative design $\alpha$ is chosen to be 1 in computing the inset of Figure 4 of the paper. The effective, design temperature, $T$, entering the Boltzmann factor of the Metropolis criterion[22], $\exp(-E_{tot}/T)$, was chosen to be $T=4$.

### Statistics of charge contacts across protein interfaces

The dataset of 122 homodimeric[6] and 48 superfamily heterodimeric ($Z>2$) (see Table 2) crystal structures was used to compute the number of *atomic* contacts across protein-protein interfaces, $n_{+-}$ (favourable (+−) contacts) and $n_{++}$ (unfavourable (++) contacts). In the experimental data analysis, $n_{+-}$ and $n_{++}$ are normalized by the total number of interface atoms. In the analysis of experimental crystal structures, we used the five atom-typing scheme [20]. In the model calculation, $n_{+-}$ and $n_{++}$ are the number of (+−) and (++) *residue* contacts, respectively, normalized by the total number of residues at the interface.

### Acknowledgements

We are grateful to S. Teichmann for communicating to us her unpublished results, and to E. Domany, K. Zeldovich, and D. Tawfik for helpful discussions. This work is supported by NIH grant GM52126.

## Figure legends

**Figure 1:** (**A**) and (**B**) snapshots represent a model superfamily heterodimer. The surfaces have identical (and random) *spatial* positions of residues, however the *identities* of residues (marked by different colours) are reshuffled within the surfaces. (**C**) Computed lowest energy distribution (LED), $P(E)$, of the interaction energy, $E$, between two model protein surfaces for random heterodimers (red line), homodimers (black line), and superfamily heterodimers (blue line). $E$ is the interaction energy per one residue in the units of $k_B T$, where $k_B$ is the Boltzmann constant. The aminoacid composition was chosen to be the composition of the homodimer surfaces data set of Table III in [6].

**Figure 2:** Toy model. There are total $Q = 16!/(8!8!) = 12870$ distinct configurations of random patterns with 8 hydrophobic residues (marked in black) on a 4x4 lattice. A contact between two hydrophobic residues reduces the energy of the system. Among these configurations, there are $Q_{\text{self}} = 8!/(4!4!) = 70$ distinct, exactly self-matching patterns, constituting strongly bound homodimers, each with 8 favourable, hydrophobic contacts (assuming a fixed mutual orientation of cubes). The probability to find a strongly bound homodimer is thus $p_{\text{homo}} \simeq 70/12870$ is ~ 70 times *larger* as compared with the probability for a strongly bound heterodimer, $p_{\text{hetero}}$.

**Figure 3:** Comparison of experimental data on the aminoacid propensities of protein interfaces and model predictions. The scatter plot of experimental versus model residue *interface propensities* for homodimers. The average compositions of residues used to generate random surfaces are taken from the homodimer data set of Bahadur *et al.*[6]

(Table III, column 5 (Surface) of Ref. [6], with surface compositions for homodimers in terms of area fraction. The resulting linear correlation coefficient between the experimental and model data is $R \simeq 0.93$. The straight line represents the linear fit to the data. Inset shows the position of the energy cut-off. The selection of strongly interacting homodimers was performed below this cut-off.

**Figure 4:** The frequency of occurrence of structurally similar (*i.e.* high Z-score) monomers within protein complexes for obligate (green bars) and transient (red bars) complexes[12] for three values of the Z-score cut-off: Z>2, Z>5, and Z>10. The monomers within all complexes have less than 25% sequence similarity. The control bars (blue bars) are computed by picking *random* pairs from *all* 3313 protein domains constituting PDUG[14], computing their Z-scores, and finally computing the corresponding frequencies of the occurrence of the high Z-score PDUG protein domain pairs. The error bars on the control represent one standard deviation.

**Figure 5:** (**A**) Real protein complexes. Computed frequency of the number of *atomic* contacts (normalized per one interface atom) across interfaces of homodimers (black bars), transient heterodimers (red bars), and obligate heterodimers (blue bars). All atomic contacts were computed regardless of atom types. Two atoms belonging to different interacting proteins chains are assumed to be in contact if they are separated by the distance of less than $7\,\mathring{A}$. The Kolmogorov-Smirnov *p*-values (comparing the similarity between the distributions) are: homodimers vs. transient heterodimers, $p \simeq 4 \times 10^{-6}$; obligate heterodimers vs. transient heterodimers, $p \simeq 10^{-4}$; and homodimers vs. obligate heterodimers, $p \simeq 0.38$. The smaller is the *p*-value, the more distinct are the two distributions. (**B**) Model protein complexes. Computed probability distribution, *P(n),* of the number of inter-surface contacts, $n$, per residue between two model protein surfaces, at the lowest (with respect to rotation) value of the inter-surface energy, $E$. Heterodimers (red line), homodimers (black line), and superfamily heterodimers (blue line). The average aminoacid compositions is from the homodimer *surface* dataset of Ref.[6].

**Figure 6:** Evolution of the Bacterial DNA-binding protein family (PFAM PF00216). The left panel is a species sub-tree with all the organisms containing a representative of this family. The right panel is a plot of distributions of pair-wise sequence identities within pairs of paralogs of monomers of this dimeric protein in a given species. Each horizontal dotted line connects the lowest and highest pair-wise sequence identity values in paralogs found in a single organism. The number of points on a line is equal to $n(n-1)/2$, where $n$ is the number of paralogs. A single point means that there were only one paralog. Absence of any points or lines indicates that there is only 1 family member in that organism.

**Figure 7:** The charge contacts frequency *differences* between superfamily heterodimeric and homodimeric protein interfaces, respectively: $P(n_{+-}^{hetero}) - P(n_{+-}^{homo})$ (filled bars), and $P(n_{++}^{hetero}) - P(n_{++}^{homo})$ (open bars). Here $n_{+-}^{hetero}$ and $n_{+-}^{homo}$ are the numbers of $(+-)$ atomic contacts (normalized by the total number of interface atoms) across superfamily hetero- and homodimeric interfaces, respectively. $n_{++}^{hetero}$ and $n_{++}^{homo}$ are the analogous numbers for $(++)$ contacts (see Methods). The results are highly statistically significant, as the computed Kolmogorov-Smirnov $p$-values demonstrate (control data not shown). Inset: The analogous data for the statistics of charge contacts across *model* protein interfaces. Model contact numbers count the residue contacts normalized by the total number of residues at the interface.

**Figure 8:** Computed positional, density-density correlation function, $g(\rho)$ (the probability distribution to find a randomly selected pair of amino acids to be separated by the distance $\rho$). The positional distribution of amino acids in strongly bound homodimers (red) has *higher* correlations as compared with weakly bound homodimers (blue). The distance is plotted in the units of the aminoacid diameter, $d_0$. The model surfaces were generated with the fixed aminoacid composition from the homodimer surface dataset[6].

## Table legends

**Table 1:** Subset of high Z-score obligate (we term these structurally similar complexes as "superfamily heterodimers") and transient complexes (selected from the entire set of complexes). Only those high Z-score complexes are chosen, where the indicated pairs of interacting chains have the sequence identity of less than 25%.

**Table 2:** Subset of 48 high Z-score ($Z>2$) obligate complexes (*i.e.*, superfamily heterodimers) used to compute the number of charge contacts across the interfaces. We included in this subset also those complexes (from the complete list of obligate complexes), which have the sequence identity between the interacting chains higher than 25%.

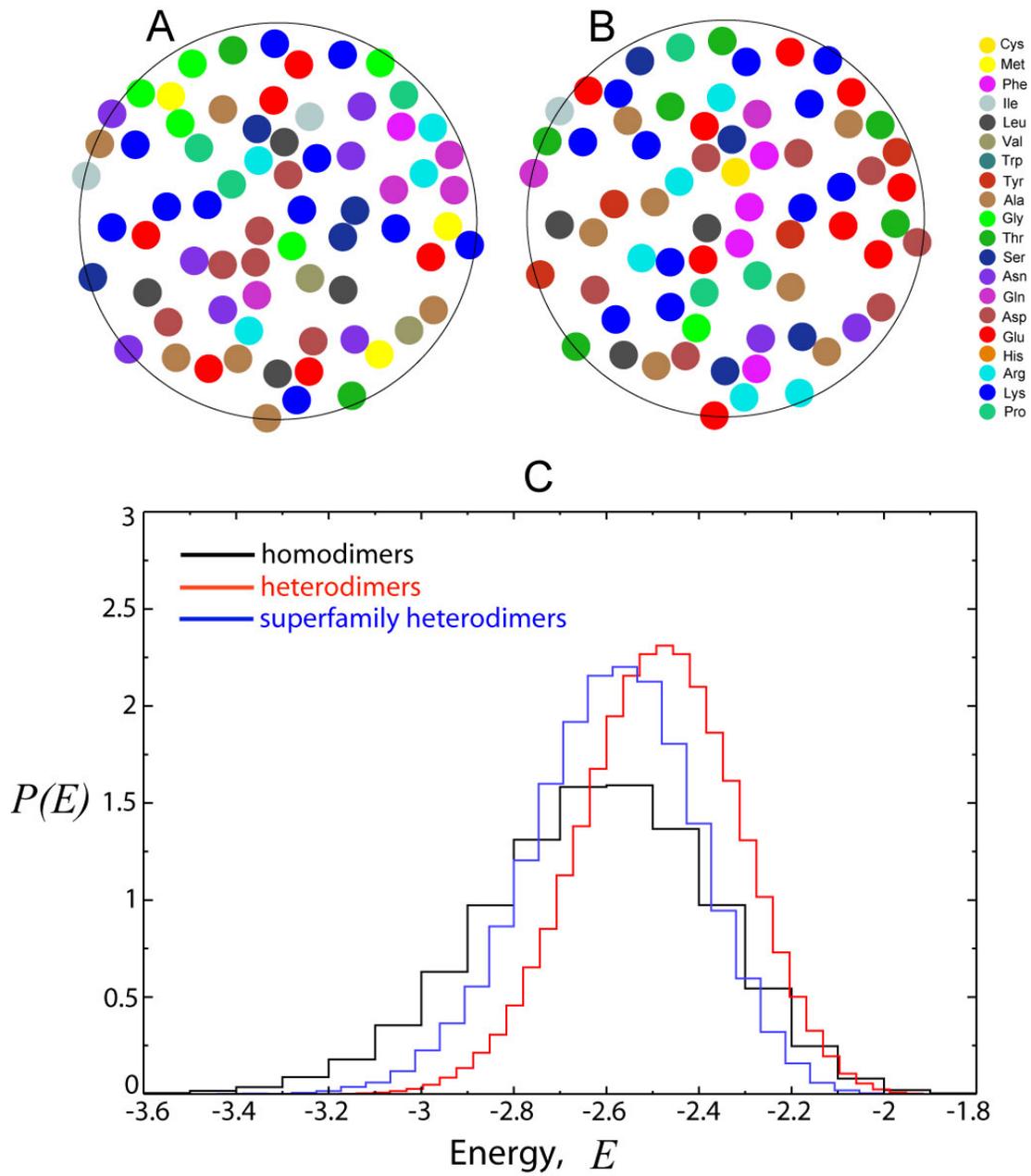

**Figure 1**

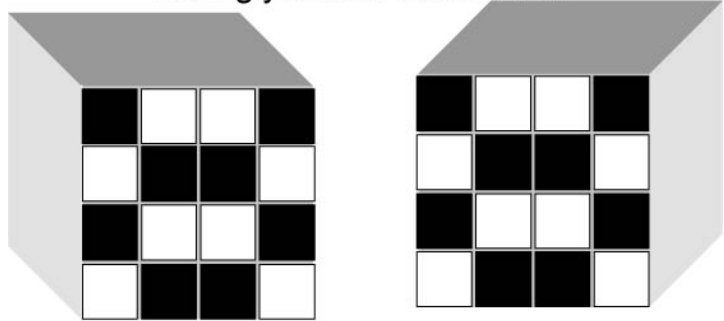

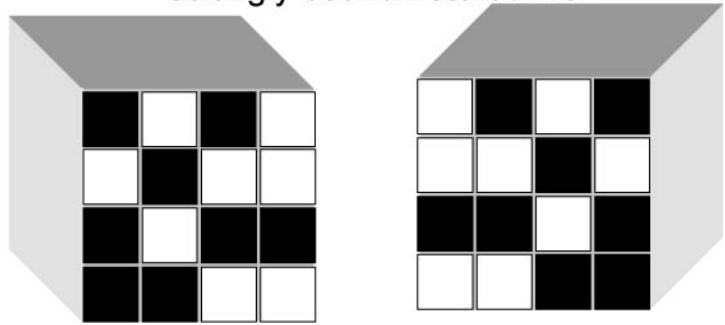

**Figure 2**

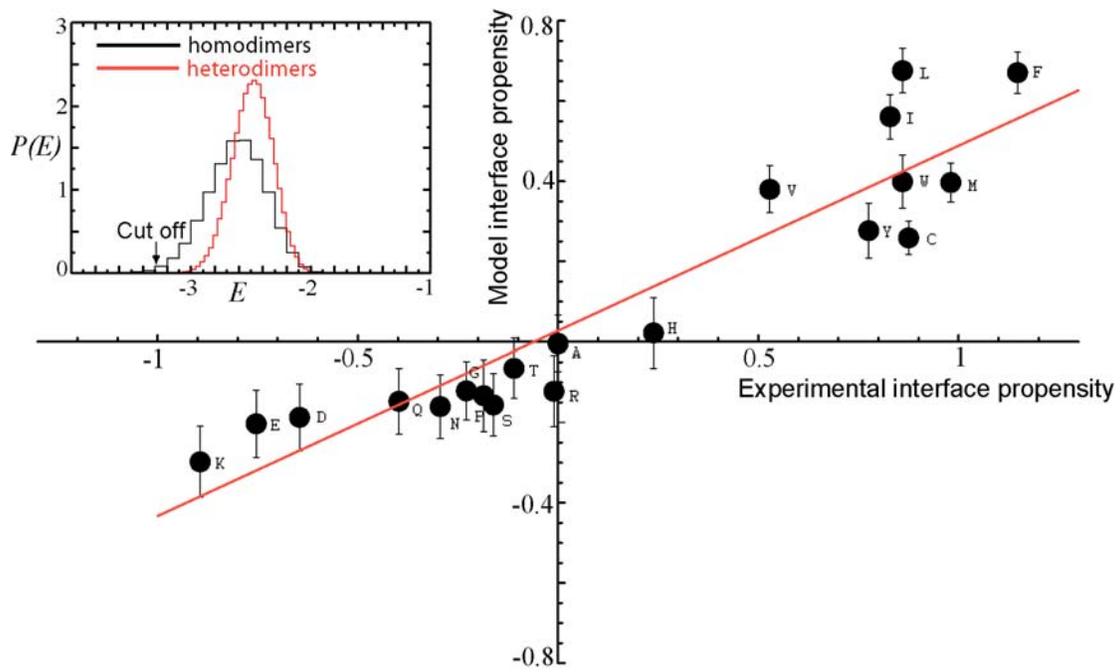

**Figure 3**

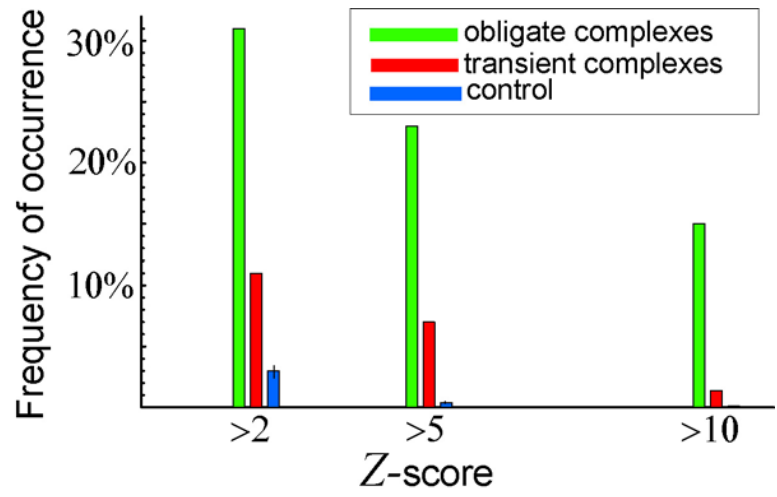

**Figure 4**

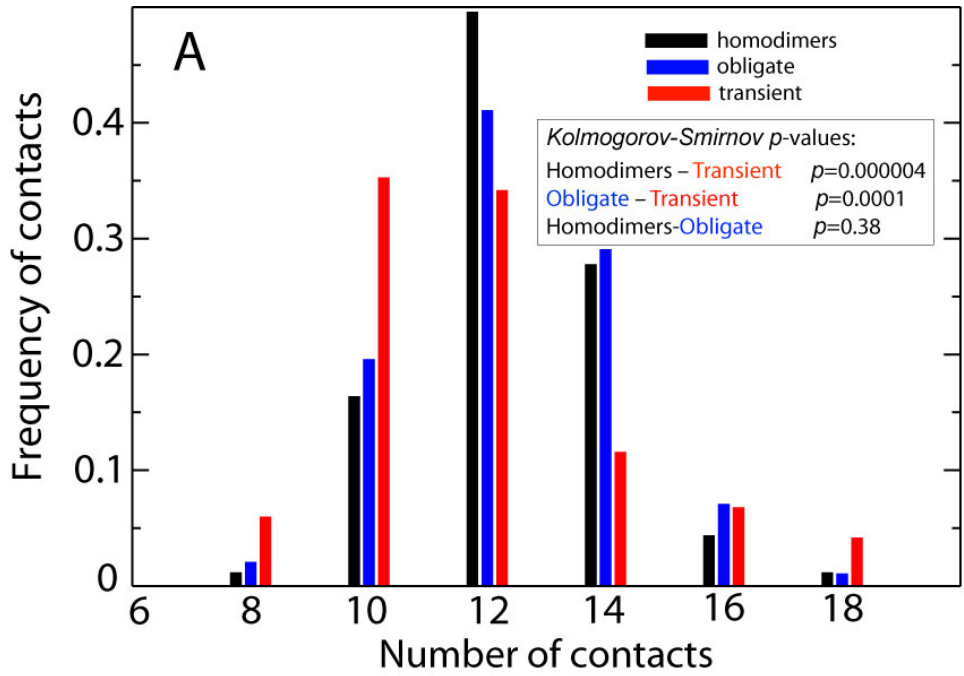

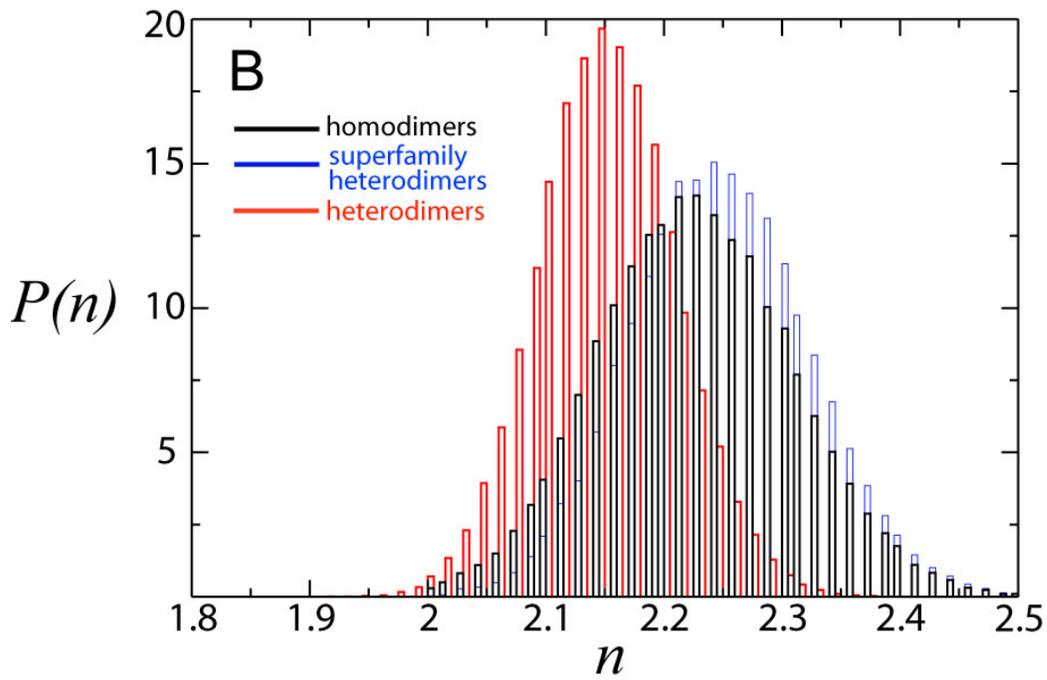

**Figure 5**

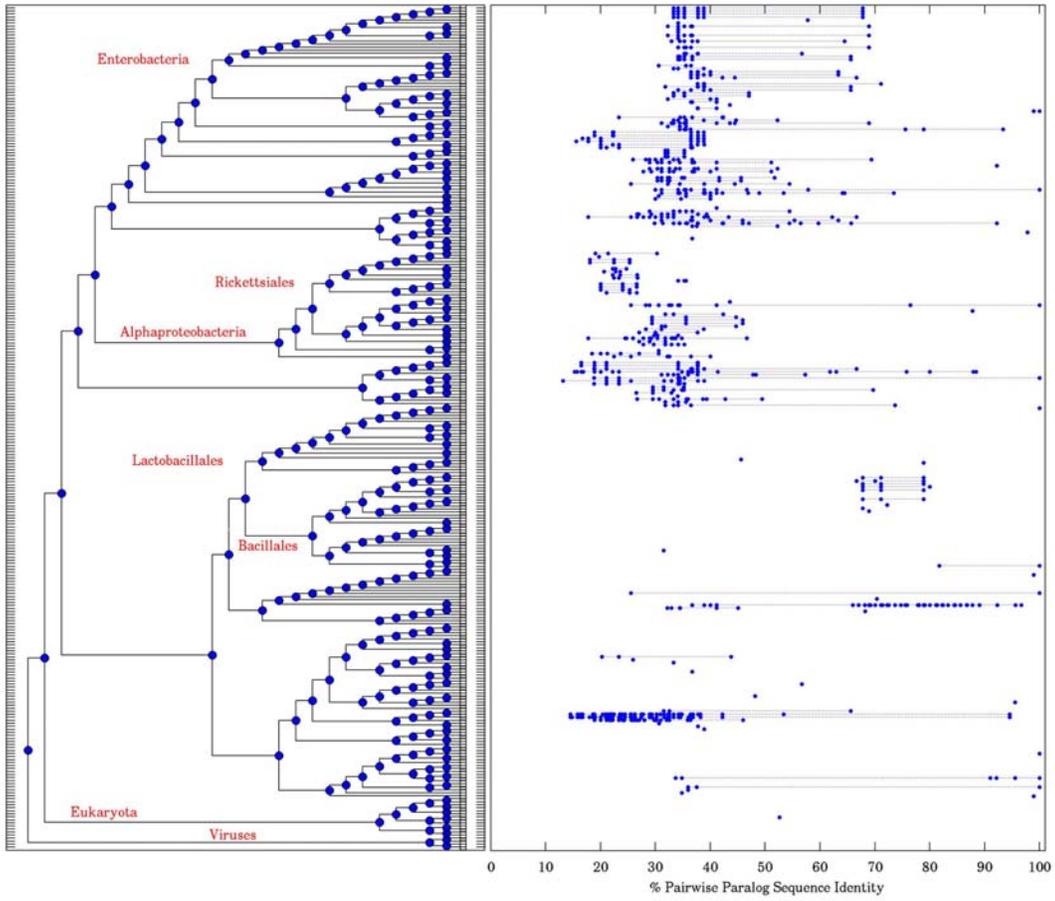

**Figure 6**

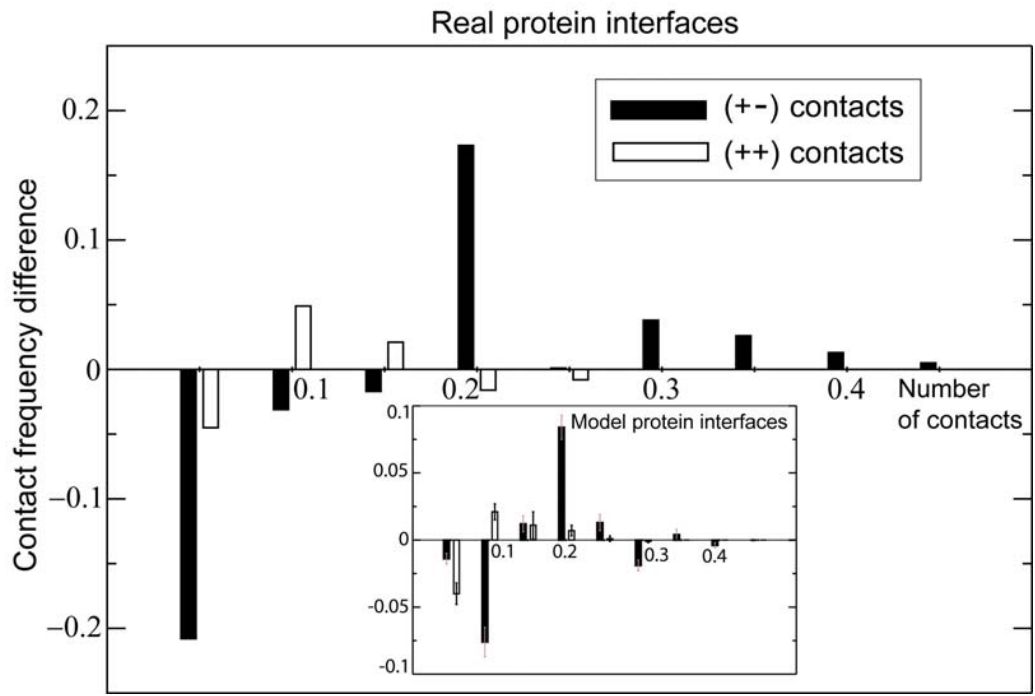

**Figure 7**

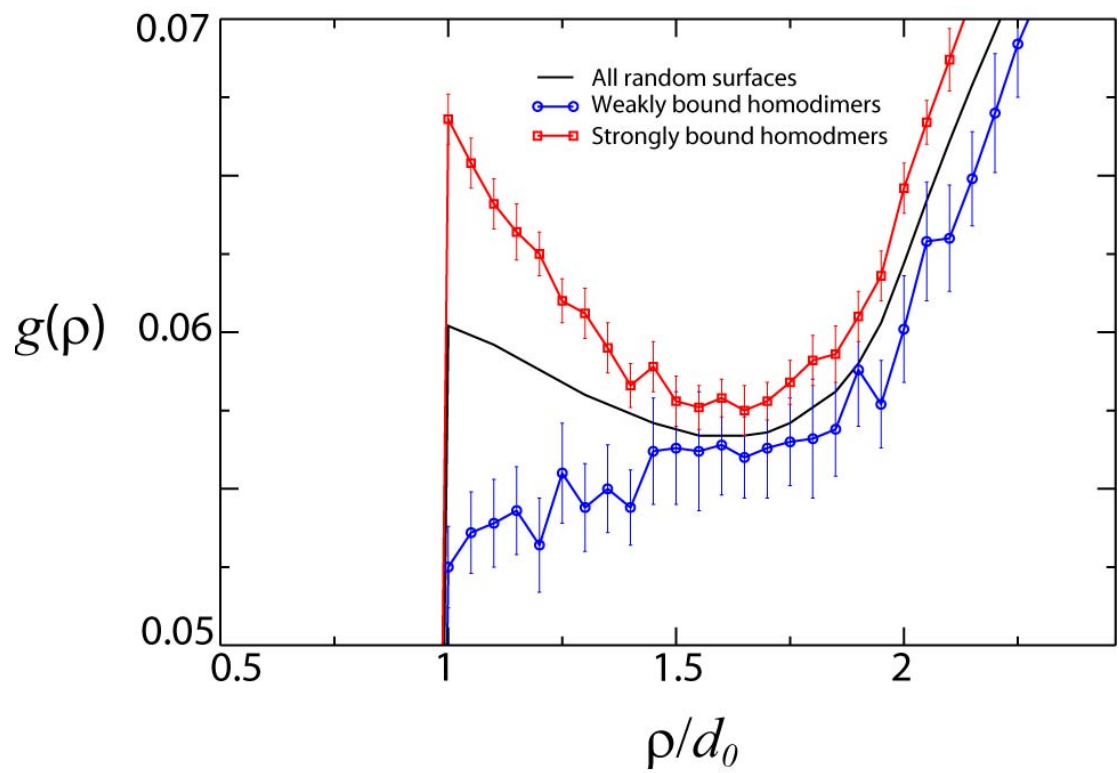

**Figure 8**

**High *Z*-score obligate complexes**

| PDB ID | *Z*-score |
|---|---|
| 1ccw A:B | 2.2 |
| 1dtw A:B | 10.3 |
| 1hsa A:B | 13.9 |
| 1e6v A:B | 23.9 |
| 1jv2 A:B | 4.6 |
| 1ep3 A:B | 2.3 |
| 1k8k D:F | 9.8 |
| 1jkj A:B | 14.0 |
| 1hzz A:C | 5.5 |
| 1dce A:B | 3.1 |
| 1efv A:B | 18.6 |
| 1b7y A:B | 16.3 |
| 1h8e A:D | 40.9 |
| 1ktd A:B | 16.0 |
| 1jmz A:B | 2.9 |
| 1jro A:B | 2.5 |
| 1hxm A:B | 17.5 |
| 1e8o A:B | 8.9 |
| 1jb7 A:B | 3.3 |
| 1hcn A:B | 5.5 |
| 1poi A:B | 6.8 |
| 1f3u A:B | 5.3 |
| 1cpc A:B | 19.2 |
| 1jk8 A:B | 16.3 |
| 1mro A:B | 24.7 |
| 1m2v A:B | 30.4 |
| 1h32 A:B | 2.2 |
| 1mjg A:M | 21.0 |
| 1ytf C:D | 7.2 |
| 2min A:B | 29.9 |

**High *Z*-score transient complexes**

| PDB ID | *Z*-score |
|---|---|
| 1dn1 A:B | 2.6 |
| 1i85 B:D | 9.2 |
| 1bqh A:G | 5.1 |
| 1ahw A:C | 5.8 |
| 1akj A:D | 5.1 |
| 1iqd A:C | 2.5 |
| 1ao7 A:D | 5.1 |
| 1iis A:C | 5.3 |
| 1im3 A:D | 6.7 |
| 1kyo O:W | 5.6 |
| 1f60 A:B | 2.6 |
| 1qav A:B | 14.2 |
| 1qo0 A:D | 5.5 |
| 1efx A:D | 5.1 |
| 1n2c A:E | 3.5 |
| 1gcq B:C | 9.5 |
| 1d2z A:B | 11.1 |
| 1m4u A:L | 4.3 |
| 1m2o A:B | 3.6 |
| 1gvn A:B | 2.0 |
| 1o94 A:C | 3.9 |
| 1i9r A:L | 2.1 |
| 1i1a B:C | 13.6 |

**Table 1**

**High *Z*-score obligate complexes used to compute the number of charge contacts across structurally similar protein interfaces**

| PDB ID | PDB ID |
|---|---|
| 1dkf A:B | 1jro A:BD |
| 1ccw A:B | 1hxm A:B |
| 1dtw A:B | 1req A:B |
| 1hsa A:B | 1kfu L:S |
| 1e6v A:B | 1jk0 A:B |
| 1jv2 A:B | 1b8m A:B |
| 1ep3 A:B | 1e8o A:B |
| 1k8k D:F | 1jb7 A:B |
| 1jkj A:B | 1hcn A:B |
| 1k8k A:B | 1poi A:B |
| 1a6d A:B | 1f3u A:B |
| 1luc A:B | 1cpc A:B |
| 1hzz AB:C | 1dxt A:B |
| 1dce A:B | 1jk8 A:B |
| 1efv A:B | 1mro A:B |
| 1ihf A:B | 1m2v A:B |
| 1b7y A:B | 1h32 A:B |
| 1gka A:B | 1mjg AB:M |
| 1fxw A:F | 3gtu A:B |
| 1h8e A:D | 1ytf BC:D |
| 1hr6 AE:B | 1spp A:B |
| 1ktd A:B | 1vkx A:B |
| 1jmz AG:B | 3pce A:M |
| 1li1 AB:C | 2min A:B |

**Table 2**